\documentclass[twocolumn]{article}
\usepackage[utf8]{inputenc}
\usepackage[english]{babel}
\usepackage{fontawesome}
\usepackage{amsmath}
\usepackage[margin=3cm]{geometry}
\usepackage{amsthm}
\usepackage{subfig}
\usepackage{amssymb}
\usepackage{stmaryrd}
\usepackage{dsfont}
\usepackage{tikz}
\usepackage{xcolor}
\usepackage{svg}
\usepackage{soulutf8}
\usepackage{booktabs}
\usepackage{marvosym}
\usepackage{xspace}
\usepackage{pgfplots}
\usepackage[hidelinks]{hyperref}
\usepackage[strings]{underscore}
\usepackage{cleveref}
\usetikzlibrary{shapes,positioning,decorations,decorations.text,arrows,decorations.pathmorphing,patterns}

\newcommand{\wrt}{with regards to\xspace}
\newcommand{\eg}{\textsl{e.g.}\xspace}

\definecolor{lviolet}{rgb}{.95,.85,1}
\definecolor{lblue}{rgb}{.85,.85,1}
\definecolor{lred}{rgb}{1,.85,.85}
\definecolor{lgreen}{rgb}{.85, 1,.85}
\definecolor{lgray}{rgb}{.92,.92,.92}
\definecolor{lorange}{rgb}{1,.92,.8}

\theoremstyle{definition}
\newtheorem{definition}{Definition}
\newcommand\blfootnote[1]{%
	\begingroup
	\renewcommand\thefootnote{}\footnote{#1}%
	\addtocounter{footnote}{-1}%
	\endgroup
}
\begin{document}
\title{MCTS-based Automated Negotiation Agent}
%
%
\author{
	Cédric L.R. Buron\thanks{Thales Research \& Technology, Palaiseau France \url{cedric.buron@thalesgroup.com}
		},
	Zahia Guessoum\thanks{LIP6, Sorbonne Université, Paris, France}\thanks{CReSTIC, Université de Reims Champagne Ardennes, France
		},
	Sylvain Ductor\thanks{Universidade Estadual do Ceará} 
	}
\maketitle              
\begin{abstract}
\blfootnote{Cite as Cédric L. R. Buron, Zahia Guessoum, and Sylvain
	Ductor. ``MCTS-based Automated Negotiation Agent''. \textsl{In The 22nd International Conference on Principles and Practice of Multi-Agent Systems (PRIMA2019)}, Turin, 2019}This paper introduces a new negotiating agent model for automated negotiation. We focus on applications without time pressure with multidimensional negotiation on both continuous and discrete domains. The agent bidding strategy relies on Monte Carlo Tree Search, which is a trendy method since it has been used with success on games with high branching factor such as Go. It also exploits opponent modeling techniques thanks to Gaussian process regression and Bayesian learning. Evaluation is done by confronting the existing agents that are able to negotiate in such context: Random Walker, Tit-for-tat and Nice Tit-for-Tat. None of those agents succeeds in beating our agent. Also, the modular and adaptive nature of our approach is a huge advantage when it comes to optimize it in specific applicative contexts.

\textbf{keywords}: Automated negotiation,
	MCTS,
	Supply Chain
\end{abstract}
	
	\section{Introduction}
	\label{sec:introduction}
	Negotiation is a form of interaction in which a group of agents with conflicting interests and a desire to cooperate try to reach a mutually acceptable agreement on an object of negotiation \cite{Baarslag2015Learning}. The agents explore solutions according to a predetermined protocol in order to find an acceptable agreement.	
	Being widely used in economic domains and with the rise of e-commerce applications, the question of automating negotiation has gained a lot of interest in the field of artificial intelligence and multi-agent systems.
	
	Many negotiation frameworks have been proposed \cite{Guttman1998Agent}. They may be characterized along different aspects, whether concerning the set of participants (\eg bilateral or multilateral), agent preferences (\eg linear or not), issues of negotiated objects (\eg discrete or continuous), or even the characteristics of the interaction protocol (\eg globally bounded in time or number of rounds). They run negotiating agents that use strategies to evaluate the received information and make proposals. Several strategies have been proposed. Either fixed or adaptive, most of them  rely on a known deadline (either in time or in rounds). However in several applications, the deadline of an agent may change over the negotiation. The negotiation horizon may vary depending on external elements such as other opportunities. To the best of our knowledge, these elements have not been taken into account so far.

	In this paper, we propose to handle this issue by designing a loosely constrained adaptive strategy for automated negotiation. This strategy considers that:  
	1) the agent preferences are nonlinear, 2) the issues of negotiated objects can be both continuous and discrete and 3) the time pressure is undefined, and therefore the deadline of the negotiation. To cope with this objective, our agent is based on General Game Playing  \cite{Finnsson2012} and Machine Learning \cite{Baarslag2015Learning}. Its strategy relies on both Monte Carlo Tree Search (MCTS), a heuristic technique that has been used with success for many kinds of games (see for instance \cite{Browne2012survey,Silver2016}), and opponent modeling techniques in order to be more efficient. 
	
    The paper is organized as follows. \Cref{application} describes our targeted industrial application. \Cref{relatedworks} provides some background on automated negotiation and AI strategies for games. \Cref{negogame} gives the theoretical and formal setting for bargaining in order to motivate the use of AI for games. \Cref{secmocana} introduces our strategy. \Cref{experimentalAnalysis} gives some details on the agent implementation and shows its performance against a Random Walker agent, Tit-for-Tat agent and Nice Tit-for-Tat agent. The last section gives concluding remarks and perspectives.
	
	\section{Target application}
	\label{application}
	
	This work is part of an industrial project that addresses an economic application, the {\it factoring}. This application requires a solution that complies with the specific scope we consider here and which is neglected by the literature.

	When a company sells goods or services to another company, it produces an invoice. The selling company is called {\it supplier} and the customer is called {\it debtor}. Each country may define a legal payment term of generally several weeks. Moreover, the principal may not pay within this payment term. In the supply chain, the debtor is often much larger than the supplier. It can therefore impose its own conditions at the expense of the supplier. The consequences are quite harmful for the latter: during those payment delays, its {\it working capital} is reduced and hence its capacity to produce, fulfill future orders or pay its own suppliers.
	
	Factoring is an interesting answer to this issue. A funding company (called a {\it factor}) accepts to fund the invoices of the supplier, by paying them immediately less than their nominal amount and assuming the delay of payment of the principal.  From the factor perspective -- generally a bank or an investment fund -- the principal can be seen as a short-term investment, where the risk  of defaulting on payment depends on the reputation of the principal. Since we consider the case where the principal is much larger than the supplier, this risk is lower and the rate is more affordable for the supplier.

	As this kind of funding may be recurrent, there is a strong interest in automating the negotiation between the supplier and the factor. Moreover, some recent works have shown an increase in the number of factoring marketplaces all around the world \cite{dziuba2018crowdfunding}. However, there are several specificities to this setting.	 
	The first specificity is the negotiation domain. Several elements are negotiated at the same time: the nominal amount to be funded and the discount rate are the primary elements of the negotiation. Also, for identical nominal amounts, a factor may ask the supplier to sell invoices of certain principals it trusts. Finally, when several invoices are available, the expected number of financing days may also be negotiated. Therefore, we consider complex issues that combine at the same time elements of various kinds:  continuous (the discount rate), numeric (the numerical amount and the financing days), and categorical (the principal). 
	
	The second specificity is related to uncertainty and resource availability. Automated negotiation often considers a deadline,  which defines the time allocated to the negotiation. Most of the negotiation strategies rely on this time pressure to compute a concession rate. In our application, time pressure is not constant over the negotiation. For the factor, the time pressure depends both on the money it has to invest and on the investment opportunities. If the factor has a lot of money and  few opportunities, the time pressure increases: the factor sees not invested money as a loss. On the contrary, when resources are limited and opportunities are common, the factor tries to get a better discount rate and the time pressure decreases. For the supplier, the situation is even more unpredictable. Time pressure depends on its opportunities to get new credit lines (including bank loans) and even on the time the principal takes to pay it: the negotiation may be brutally interrupted at some point if the payment of an invoice makes it useless for the supplier.
	
	Automated negotiation components
 relying either on a deadline commonly known by the agents, or on a deadline private to each of them is therefore not applicable to our target applications.	
	In our target application the negotiation domain consists of numerical, continuous and categorical issues, the agent preferences are nonlinear, and the time pressure is dynamic.
	
	\section{Related work}
	\label{relatedworks}

    Our agent is at the meeting point of automated negotiation and Monte Carlo methods applied to games. In this section, we introduce both domains.
	
	\subsection{Automated Negotiation}
	
	Various authors have explored the negotiation strategies with different perspectives. Most of those works have identified three components that make up the ``BOA'' architecture \cite{Baarslag2016Exploring}: a \textbf{Bidding strategy} defines the offers the agent sends to its opponent, an \textbf{Acceptance strategy} defines whether the agent accepts the offer it just received or if it makes a counterproposal, and an \textbf{Opponent modeling}  models some features of the opponents, as its bidding strategy, its preference domain and its acceptance strategy. The latter aims to improve the efficiency of the bidding strategy and/or the acceptance strategy of the agent.
	The following subsections present the related work for each component.
	
	\subsubsection{Bidding Strategy}
	
	Bidding strategies may depend on several elements: the history, including the concessions made by the opponent and/or a negotiation deadline, the utility function of the agent, and the opponent model. 
	Faratin \textsl{et al.} proposed the so-called tactics that mainly rely on the criticality of the resources, the remaining time before the deadline is reached or the concessions made by the opponent. All of them except the last rely on a known deadline. The latter has been extended to create more complex strategies, as the Nice tit-For-Tat agent \cite{Baarslag2013tit}, which uses learning techniques in order to improve it. Genetic algorithms \cite{Jonge2016} use generated proposals as individuals. They rely on the time pressure  for the variation of their proposals and make some that are acceptable for their opponent.
	
	This paper advances the state of the art in the bidding strategies by introducing and evaluating a strategy that considers the negotiation as a game and uses the very efficient Monte Carlo Methods.
	
	\subsubsection{Acceptance Strategy}
	Acceptance strategies can be divided into two main categories \cite{Baarslag2013}. The first category is called ``myopic strategies'' as they only consider the last bid of the opponent. An agent may accept an offer when (1) it is better than the new one produced by its own bidding strategy, (2) it is better than the last one made by the agent,  (3) it is above a predetermined threshold, or (4) it embodies any combination of the previous ones. The second category consists of ``optimal strategies'' \cite{Baarslag2013} that rely on an opponent bidding strategy in order to optimize the expected utility. They are based on the concessions made by the opponent and a prediction that the expected utility of the agent should increase while the deadline is getting closer. So, the first category is not suitable for our context. We therefore propose to use the second category in our agent model. 
	
  	\subsubsection{Opponent modeling}
	\cite{Baarslag2015Learning} presents an exhaustive review of the opponent modeling techniques related to automated negotiation. 
    They are generally used to model (1) opponent bidding strategy, (2) its utility and (3) its acceptance strategy as well as private deadline and a reserve price, depending on which of these elements are relevant in each context.
	
	There are two main methods to model adaptive \textbf{bidding strategy} which does not rely on the deadline: neural networks and time series-based techniques. Neural networks use a fixed number of previous offers as input, and the expected value for the next proposal as output. Time series methods can be generalized very easily. Among them, the Gaussian process regression is a stochastic technique which has been used with success by \cite{Williams2011Using}.
	Due to its nature, it can generate various proposals at each negotiation turn. Those proposals are proportional to a likelihood provided by the regression. We select this technique to model the opponent bidding strategy since it is particularly adapted to the Monte Carlo Tree Search.
	
	The opponent \textbf{utility} is generally considered as the weighted average of partial utility functions for each issue. Two families of methods are used to model them. The first one is based on the frequency of each value among the previous opponent bids. The methods of this family make the hypothesis that the most frequent values are the ones the opponent prefers, and that the most stable issues are the most important for it. They are relevant in the cases where the negotiation domain only consists of discrete issues; their extension to the continuous case is not suitable to complex domains, as it requires the definition of a distance function which depends on the negotiation domain. The second family of methods is based on Bayesian Learning \cite{Hindriks2008Opponent}. It is well suited for the continuous case and can be easily extended to categorical domains. We use it as is for the numerical issues and make an extension for the categorical ones. The latter is presented in \cref{ch5-2-opp-modeling-utility}.
	
	Their is two ways to learn the opponent \textbf{acceptance strategy}: either by assuming that the opponent has a myopic strategy or by using neural networks \cite{Fang2008Opponent}. The latter is quite expensive in terms of computation time. The weights of the network must be updated each time the opponent makes a new proposal.
	
	\subsection{Monte Carlo Tree Search}
	
	Monte Carlo methods are regularly used as heuristics for games. Rémi Coulom \cite{coulom2006efficient} proposes a method to combine the construction of a game tree -- a traditional method for games that has proved to be very to be effective -- with Monte Carlo techniques. This method is called Monte Carlo Tree Search (MCTS) and it has been improved using various extensions \cite{Browne2012survey}, including pruning the less promising branches of the tree. It has met great success in games, particularly games with high branching factor \cite{Silver2016}.
	
	MCTS consists of 4 steps. A \textbf{selection} is dedicated to the exploration of the already built part of the tree, based on a predefined strategy. While exploring a node, the algorithm chooses whether to explore a lower-level branch or expand a new branch. In the latter case, there is an \textbf{expansion} of a new node; it is created just below the last explored one. Once a new node has been expanded, a \textbf{simulation} of the game is performed until a final state is reached. Finally, outcomes of the final state are computed and a \textbf{backpropagation} of them is made over all the nodes that have been explored.
	
	\cite{Jonge2017Automated} presents a recent attempt to exploit MCTS for  General Game Playing, with Automated Negotiation as a potential application. The negotiation domain considered is limited to a single-issue, discrete domain with complete information  (each agent knows what is the optimal deal for its opponent). Our work is specifically made for recent evolution of Automated Negotiation, focusing thus on multi-issues, combining continuous, numerical and categorical domains with incomplete information (the agent has no information on the opponent utility profile). These differences have consequences on the technical aspects of these works. The referenced works use Upper Confidence Trees, inapplicable in our case since it imposes the number of possible moves at each step to be finite. Also, \cite{Jonge2017Automated}  does not require opponent modeling, since the opponent utility profile is known.

	\section{Negotiation as a game}
	
	\label{negogame}
	Monte Carlo Tree Search has been applied with success to extensive games. In this section, we show how it is possible to represent negotiation using this model. We first associate each aspect of negotiation to a game element. We then describe specificities of negotiation that prevent us from using most common MCTS selection, expansion and simulation strategies. 

	An extensive game \cite{Osborne1994} consists of a set of players, the set of all possible game histories, a function mapping each non-terminal history to the player who must play then and a preference profile. By using this definition, it is possible to define a bargaining $\mathcal{B}$ as an extensive game:
	\begin{definition}[Bargaining]
		\label{ch4-bargaining}
		A bargaining can be represented as triple $\mathcal{B} = \left(H,A,\left(u_i\right)_{i\in\llbracket 1,2\rrbracket}\right)$ where:
		\begin{enumerate}
			\item $A$ is the set of two \textbf{players}: the {\it buyer} (player $1$) and the {\it seller} (player $2$),
			\item $H$ is the set of \textbf{possible histories} of the negotiation. Each history consists of the sequence of messages the agents sent to each other: {\it proposals}, {\it acceptance} and {\it rejection} messages. Terminal histories are the histories ending by acceptance or rejection, and infinite histories. Each message is a pair $(\alpha, c)$ where $\alpha$ is the speech act (performative) of the message and $c$ is the content of the message, \textsl{i.e.} a list of couples $(k,v)$ where $k$ is the key of an issue of the negotiation domain and $v$ is the corresponding value. The history can be divided in two parts,
 each part corresponding to the messages sent by one of the agents: $h_i = (\alpha,c)_i$,
			\item the $\mathbf{player}$ function is based on the parity of the size of the history (we suppose that the buyer (player 1) always plays first). Therefore $\forall h\in H, player(h)=2-(|h| \mod 2)$ where $|h|$ is the size of $h$,
			\item the \textbf{preference profile}, $u_i, i \in \{1,2\}$, is an evaluation of terminal histories \wrt each player. If the history ends with an acceptance of the agent, $u_i$  returns the utility associated by the agent, if not it returns a specific value that may depend on the engaged resources.
		\end{enumerate}
	\end{definition}
	
	The representation of bargaining as a game has already been investigated in \cite{nash1950bargaining,rubinstein1982perfect} for single issue bargaining. More complex domains has been initially dealt by making the assumption that the agents proposals are independent of the opponent's ones \cite{Faratin1998Negotiation}. However in recent advances in automated negotiation, the agent bidding strategies are generally adaptive \textsl{i.e.} the proposals made by an agent depend on its opponent's ones.
	
	However, negotiation is not a classical combinatorial game as Chess or Go. Its resolution is a challenge to the MCTS approach for three reasons. First, it is a non-zero sum game: agents try to find a mutually beneficial agreement which is often much better for both agents than their reserve utility \textsl{i.e.} the situation where the agents do not find an agreement. Second, it is an incomplete information game: the agent preference profiles are unknown to their opponents and generally modeled by them. These two specificities make it impossible to use the most common implementation of MCTS, the Upper Confidence Tree \cite{Kocsis2006Bandit}. Last, we consider a large and complex domain that encompasses numeric, continuous and categorical issues, with nonlinear utility functions and possibly infinite game trees. This has several consequences on the way the tree is explored, in particular on the criterion followed to expand a new node.

	\section{MCTS-based agent}
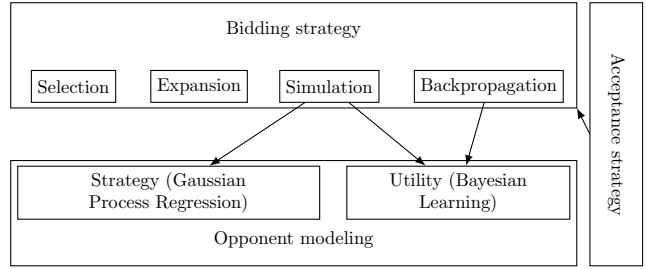
\begin{figure}[!htbp]
		\centering
		\begin{tikzpicture}[scale=.7, every node/.style={scale=.7}]
		\node at (4,3.6) [draw,  anchor=south east, minimum height=4ex] (selection) {Selection};
		\node at (6.5,3.6) [draw, anchor=south east, minimum height=4ex] (expansion) {Expansion};
		\node at (9,3.6) [draw, anchor=south east, minimum height=4ex] (simulation) {Simulation};
		\node at (12.5,3.6) [draw, anchor=south east, minimum height=4ex] (backprop) {Backpropagation};
		
		\node at (7.375,5) {Bidding strategy};
		
		\draw (2,5.5) -- (12.75,5.5) -- (12.75,3.5) -- (2,3.5) -- cycle;
		
		\node at (5,2.4) [draw, text width=5.5cm, text centered, anchor=north] (strategy) {Strategy (Gaussian Process Regression)};
		\node at (10.5,2.4) [draw, text width=4cm, text centered, anchor=north] (utility) {Utility (Bayesian Learning)};
		
		\node at (7.375,1) {Opponent modeling};
		
		\draw (2,2.5) -- (12.75,2.5) -- (12.75,.5) -- (2,.5) -- cycle;
		
		\node at (13.5,3) [text width=5cm, text centered, rotate=-90] {Acceptance strategy};
		\draw (13,5.5) -- (13,.5) -- (14, .5) -- (14,5.5) -- cycle;
		
		\draw [->, >=latex] (simulation) -- (strategy);
		\draw [->, >=latex] (simulation) -- (utility);
		\draw [->, >=latex] (backprop) -- (utility);
		\draw [->, >=latex] (13,3) -- (12.75,3.5);
		
		\end{tikzpicture}
		\caption{Interaction between the modules of our agent}
		\label{interaction}
	\end{figure}

	As we explained in the previous section, negotiation is a particular game. It is therefore required to adapt the heuristics traditionally used for games to its specificities. In this section, we present our automated negotiation agent relying on MCTS. The agent architecture is composed of three modules presented in \Cref{interaction}. The bidding strategy module implements MCTS and uses the opponent modeling module. The latter consists of two submodules: one models the opponent utility, the other models its bidding strategy. The last module is the acceptance strategy, which makes a comparison between the last proposal from the opponent and the bid generated by the bidding strategy. Each of the agent submodules and their interactions are described in this section.
	\label{secmocana}
	
	\subsection{Opponent modeling}
	\label{ch5-2-opponent-modeling}
	
	In order to improve the efficiency of MCTS, we model both the bidding strategy and the utility of the opponent.
	
	\subsubsection{Bidding Strategy Modeling}
	\label{ch5-2-Learn}
	The goal of this model is to predict what proposal the opponent will make at turn $x_*$. To do so, we use Gaussian Process Regression \cite{Rasmussen2006Gaussian}. This method produces a Gaussian with the mean corresponding to the value predicted by the algorithm and a standard deviation corresponding to the uncertainty induced by the model.
	
	The first step is to compute the covariance matrix $K$ which represents the proximity between the turns $(x_i)_{i\in\llbracket 1, n\rrbracket}$ of the sequence, based on a covariance function, also called kernel $k$. Let \begin{equation}\def\arraystretch{0.7}
	K= \left( \begin{array}{ccc}
	k(x_1, x_1) & \dots & k(x_1, x_n) \\
	\vdots &  & \vdots \\
	k(x_n, x_1) & \dots & k(x_n, x_n) \end{array} \right)
	\end{equation}
	
	we then compute the distance between the turn of the predicted proposal $x_*$ and the previous turns in the vector $K_*$:
	
	\begin{equation}K_* = \left(k(x_*, x_1), \dots, k(x_*, x_n)\right)\end{equation}
	
	The Gaussian process regression relies on the supposition that all these values are the dimensions of a multivariate Gaussian. Using results on multivariate Gaussian, we can compute
	\begin{equation}
	\overline{y_*} = K_*K^{-1}\mathbf{y}
	\end{equation}
	\begin{equation}
	\sigma^2_* = \mathrm{Var}(y_*) = K_{**}-K_*K^{-1}K_*^\top
	\end{equation}
	where $K_{**} =k(x_*, x_*)$. The result corresponds to a Gaussian random variable with mean $\overline{y_*}$ and standard deviation $\sigma_*$.
	
	One of the capital aspects of Gaussian process regression is the choice of the kernel. The most common ones are radial basis functions (RBF), rational quadratic functions (RQF), Matérn kernel and exponential sine squared (ESS). These kernels are used to define distance between the turns of the bargaining. We tested these four kernels on various negotiations among finalists of ANAC 2014 (bilateral general-purpose negotiation with nonlinear utilities). The \cref{kernel-table} shows the results of GPR for each of the aforementioned kernels. We generated randomly 25 negotiation sessions and modeled two agents using each kernel. We got a total of 50 models by kernel. Each bid of each sequence is predicted using previous proposals and is used to predict following ones. The table shows average Euclidean distance between the actual proposals and the predicted sequences. The lower the value, the closer is the prediction to the actual proposal.
	The kernel that got the best result is the Rational Quadratic Function. Our agent therefore uses this one.
	
	\begin{table}[!ht]
		\setlength{\tabcolsep}{12pt}	
		\centering
		\begin{tabular}{l | c c c c}
			\textsl{Kernel} & RBF & RQF & Matérn & ESS \\\hline
			avg. distance & 43.288 & \textbf{17.766} & 43.228 & 22.292
		\end{tabular}
		\caption{Average Euclidean distance between actual proposals of a bargaining and values predicted by GPR, depending on the used kernel.}
		\label{kernel-table}
	\end{table}
	
	This method  also allows  to make predictions on the categorical issues. The method presented in chapter 3 of \cite{Rasmussen2006Gaussian} is also possible and relies on Monte Carlo method for some integration estimation.
	
	\subsubsection{Preference Profile Modeling}
	\label{ch5-2-opp-modeling-utility}
	
	Bayesian learning described in \cite{Hindriks2008Opponent} makes the only
 supposition that an agent makes concessions at roughly constant rate. Though this constraint may seem tough, it is relatively low in comparison with other methods.
	
	The opponent utility is approximated by a weighted sum of triangular functions. A function $t$ of $[a,b]\subset\mathds{R}$ in $[0,1]$ is called triangular if and only if:\begin{itemize}
		\item $t$ is linear and either $t(a)=0$ and $t(b)=1$ or $t(a)=1$ et $t(b)=0$, or:
		\item there is some $c$ in $[a,b]$ such that $t$ is linear on $[a,c]$ and $[c,b]$, $t(a)=0$, $t(b)=0$ and $t(c)=1$.
	\end{itemize}
	The method can be divided into two steps. First, the agent generates a predetermined number of hypotheses on the utility function. These hypotheses are composed of weighted sums of triangular functions (one per issue). Each issue is therefore associated with a weight and a triangular function.
	
	The estimated utility of the opponent is the weighted sum of these hypotheses where each weight is the probability computed using Bayesian learning. This method does not make any supposition on the opponent strategy, which makes it more general than the frequency based techniques.
	
	This method can be naturally extended to the categorical issues. Given a categorical issue $C=\{C_1,\dots,C_n\}$, the partial utility function is chosen among the set of functions of $[0,1]^C$.
	
	\subsubsection{Acceptance Strategy Model}
	Acceptance strategy is modeled in a very simple way: a simulated agent accepts the proposal from its opponent if and only if its utility is better than the utility generated by the bidding strategy model. This method presented by Baarslag \textsl{et al.} in \cite{Baarslag2015Learning} is not computationally expensive.
	
	\subsection{MCTS-Based Bidding Strategy}
	Monte Carlo methods are very adaptive, and achieve promising results in various games, including games with high branching factor. In this section, we describe the way we have  adapted these methods to the negotiation context.
	
	\subsubsection{Raw MCTS}
	As we explained before, MCTS is a general algorithm and relies on several strategies. Each time an agent needs to take a decision, it generates a new tree and explores it using MCTS. The most common implementation of MCTS is Upper Confidence Tree (UCT). This method has proved to be very efficient, particularly for Go. Nevertheless, it expands a new node whenever it explores a node whose children have not all been explored, which is not applicable when issues are continuous. Beyond that, in the case where the branching factor is not small in comparison with the number of simulation, UCT keeps expanding a new node without exploring deeper nodes, which loses the interest of MCTS compared to flat Monte Carlo. In our context, it is therefore necessary to define a different implementation of MCTS:
	\begin{description}
		\item[Selection] For this step, we use progressive widening, as described by \cite{Couetoux2013Monte}. The expansion criterion of the progressive widening states that a new node is expanded if and only if:
		\begin{equation}
		n_p^\alpha\geq n_c
		\label{prog_wid_equation1}
		\end{equation}
		where $n_p$ is the number of times the parent has been simulated, $n_c$ is its number of children and $\alpha$ is a parameter of the model. If the result is that a new node is not expanded, the selected node is the node $i$ maximizing:\begin{equation}
		W_i = \dfrac{s_i}{n_i+1} + C\times n^\alpha\sqrt{\frac{\ln(n)}{n_i+1}}
		\label{prog_wid_equation2}
		\end{equation}
		where $n$ is the total number of simulations of the tree, $s_i$ is the score of the node $i$ and $C$ is also a parameter of the model.
		\item[Expansion] The content of the expanded node is chosen randomly among all the possible bids of the domain with an even distribution
		\item[Simulation] During the simulation, the model of the opponent bidding strategy is used in order to make the simulation more representative. The model of the opponent utility and its acceptance strategy are used to decide when it accepts a proposal.
		\item[Backpropagation] The backpropagation also uses the opponent utility model. The utility of both agents is computed and the scores of both agents is updated for each visited node. The utility of the agent itself is computed using its real preference profile.
	\end{description}
	
	\subsubsection{Pruning}
	\label{ch5-mcts-pruning}
	In order to explore only the interesting nodes for the agent, it is possible to use the knowledge of the agent on the game to prune the less promising branches of the tree. Though we do not have much information in our context, we decide that the opponent should find any offer it made acceptable. We therefore decide to prune all the branches of the tree where our agent makes a proposal less interesting than the best proposal received from the opponent: the goal is to use the best proposal of the opponent as a lower bound and to try to improve on this basic value.

	\section{Experiments}
	\label{experimentalAnalysis}
	\label{xp}
	In order to evaluate our agent, we use the \textsc{Genius} \cite{Lin2014} framework. We confront our agent with the only agents that do not require a deadline for their strategy to the best of our knowledge: two variants of the Tit-for-Tat agent and a RandomWalker agent. In this section, we describe the implementations of our agent and those three. We then describe in detail the experimental protocol and finally we present the achieved results. 
	
	\subsection{Implementation}
	
	Our agent is developed in Java and consists of a set of independent modules connected to \textsc{Genius} platform thanks to interfaces. \Cref{architecture} presents this architecture (the \faEnvelope{} represents the messages sent and received by the agent). It consists of  the components of  \Cref{secmocana}: a module for the MCTS-based bidding strategy (with classes for the Monte Carlo Tree, its nodes, etc.), a module for opponent utility modeling and another one for the opponent bidding strategy modeling. We use the same  acceptance strategy for both our agent and the model it uses for its opponent. As they are very simple, there is no module dedicated to them.
	
	As MCTS are computationally expensive and take quite a long time, we parallelized them. The stochastic aspect of the opponent bidding strategy model ensures that two simulations going from the same branch of the tree are not similar, while still prioritizing the most probable values. The Gaussian process regression, which relies on matrix computation, has been developed using the Jama library\footnote{\url{http://math.nist.gov/javanumerics/}}. The parameters of the kernel are optimized using the Apache Commons Math library utilities\footnote{\url{http://commons.apache.org/proper/commons-math/}}.
	
	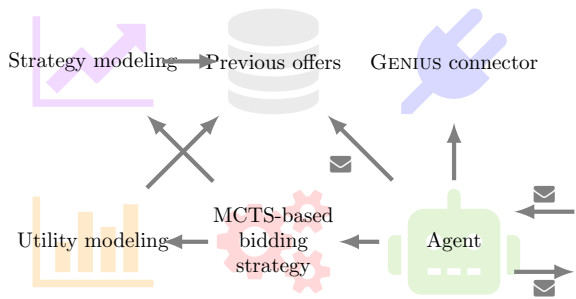
\begin{figure}[!htbp]
		\centering
		\begin{tikzpicture}[->, ultra thick, >=latex, every node/.style={scale=.8},scale=.8]
		\node at (6,0) (genius){\textcolor{lblue}{\fontsize{50pt}{5pt}\faPlug}};
		\node at (3,-3) (mcts){\textcolor{lred}{\fontsize{50pt}{5pt}\faGears}};
		\node at (6,-3) (agent){\textcolor{lgreen}{\includegraphics[height=50pt]{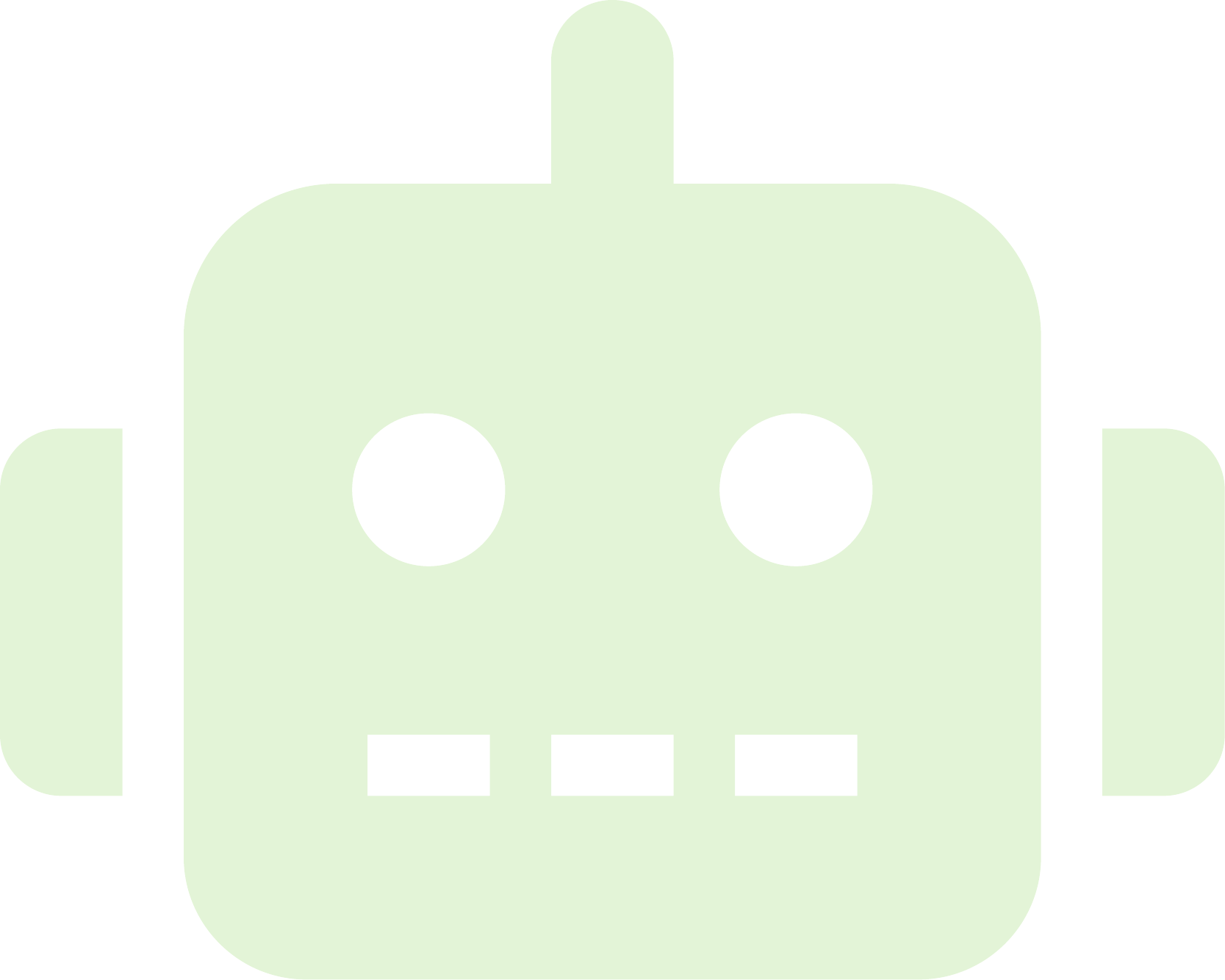}}};
		\node at (0,0) (strategy){\textcolor{lviolet}{\fontsize{50pt}{5pt}\faLineChart}};
		\node at (0,-3) (utility){\textcolor{lorange}{\fontsize{50pt}{5pt}\faBarChart}};
		\node at (3,0) (data){\textcolor{lgray}{\fontsize{50pt}{5pt}\faDatabase}};
		
		\node at (6,0) [text width=3cm, text centered]{\textsc{Genius} connector};
		\node at (3,-3) [text width=2.5cm, text centered]{MCTS-based bidding strategy};
		\node at (6,-3) [text width=3cm, text centered]{Agent};
		\node at (0,0) [text width=3cm, text centered]{Strategy modeling};
		\node at (0,-3) [text width=3cm, text centered]{Utility modeling};
		\node at (3,0) [text width=3cm, text centered]{Previous offers};
		
		\draw [gray](7,-3.5) -- (8,-3.5) node [midway, below] {\faEnvelope};
		\draw [gray](8,-2.5) -- (7,-2.5) node [midway, above] {\faEnvelope};
		\draw [gray](agent) -- (data) node [midway, below left] {\faEnvelope};
		\draw [gray](mcts) -- (strategy);
		\draw [gray](mcts) -- (utility);
		\draw [gray](agent) -- (mcts);
		\draw [gray](agent) -- (genius);
		\draw [gray](strategy) -- (data);
		\draw [gray](utility) -- (data);
		\end{tikzpicture}
		\caption{Software architecture of the modules of our agent}
		\label{architecture}
	\end{figure}
	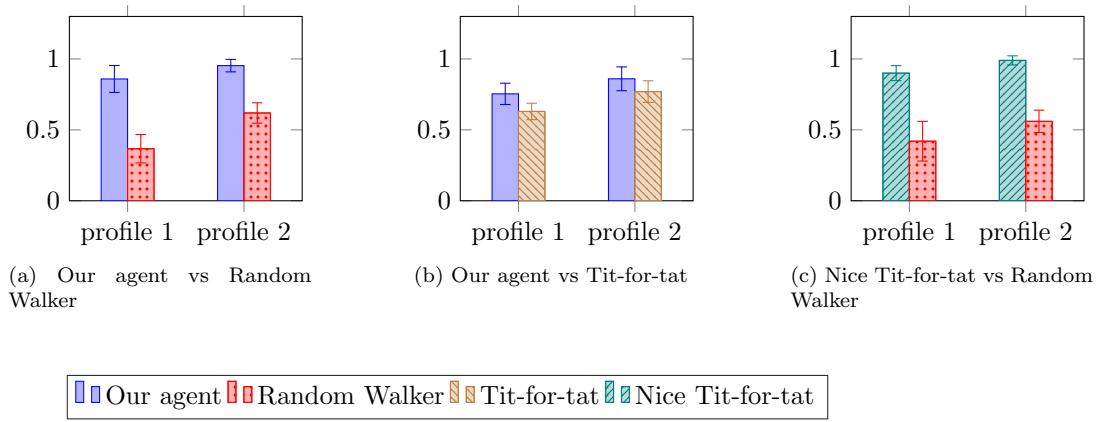
\begin{figure*}[!htb]
		\hfill
		\subfloat[Our agent vs Random Walker]{
			\begin{tikzpicture}
			\begin{axis}[
			nodes/.style={color=black},
			color=black,
			ymin=0,
			ymax=1.3,
			ybar=0pt,
			enlarge x limits=0.5,
			xtick={1,2},
			xticklabels={profile 1,profile 2},
			width=.3\textwidth
			]
			\addplot+[style={color=blue,fill=blue!30},error bars/.cd,
			y dir=both,y explicit]
			coordinates {
				(1,0.859) +- (0.0, 0.095)
				(2,0.953) +- (0.0, 0.044)};
			\addplot+[style={color=red,fill=red!30,postaction={pattern=dots,pattern color = red}},error bars/.cd,
			y dir=both,y explicit]
			coordinates {
				(1,0.367) +- (0.0, 0.1)
				(2,0.619) +- (0.0, 0.072)};
			\end{axis}
			\end{tikzpicture}
			\label{res-rw}
		}\hfill
		\subfloat[Our agent vs Tit-for-tat]{
			\begin{tikzpicture}
			\begin{axis}[
			nodes/.style={color=black},
			color=black,
			ymin=0,
			ymax=1.3,
			ybar=0pt,
			enlarge x limits=0.5,
			bar width=10,
			xtick={1,2},
			xticklabels={profile 1,profile 2},
			width=.3\textwidth
			]
			\addplot+[style={color=blue,fill=blue!30},error bars/.cd,
			y dir=both,y explicit]
			coordinates {
				(1,0.754) +- (0.0, 0.075)
				(2,0.86) +- (0.0, 0.084)};
			\addplot+[style={color=brown,fill=brown!30,postaction={pattern=north west lines,pattern color = brown}},error bars/.cd,
			y dir=both,y explicit]
			coordinates {
				(1,0.63) +- (0.0, 0.058)
				(2,0.77) +- (0.0, 0.076)};
			\end{axis}
			\label{res-tft}
			\end{tikzpicture}
		}\hfill
		\subfloat[Nice Tit-for-tat vs Random Walker]{
			\begin{tikzpicture}
			\begin{axis}[
			nodes/.style={color=black},
			color=black,
			ymin=0,
			ymax=1.3,
			ybar=0pt,
			enlarge x limits=0.5,
			bar
 width=10,
			xtick={1,2},
			xticklabels={profile 1,profile 2},
			width=.3\textwidth
			]
			\addplot+[style={color=teal,fill=teal!30,postaction={pattern=north east lines,pattern color = teal}},error bars/.cd,
			y dir=both,y explicit]
			coordinates {
				(1,0.90) +- (0.0, 0.053)
				(2,0.99) +- (0.0, 0.032)};
			\addplot+[style={color=red,fill=red!30,postaction={pattern=dots,pattern color = red}},error bars/.cd,
			y dir=both,y explicit]
			coordinates {
				(1,0.42) +- (0.0, 0.14)
				(2,0.56) +- (0.0, 0.079)};
			\end{axis}
			\end{tikzpicture}
			\label{res-ntft}
		}\hfill\\\hfill
		\subfloat{\begin{tikzpicture} 
			\begin{axis}[%
			height=2cm,
			width=\textwidth,
			hide axis,
			xmin=10,
			xmax=50,
			ymin=0,
			ymax=0.4,
			ybar,
			legend columns=4,
			legend style={draw=white!15!black,legend cell align=left,
			at={(0.5,0)},anchor=north
			}
			]
			\addlegendimage{color=blue,fill=blue!30}
			\addlegendentry{Our agent};
			\addlegendimage{color=red,fill=red!30,postaction={pattern=dots,pattern color = red}}
			\addlegendentry{Random Walker};
			\addlegendimage{color=brown,fill=brown!30,postaction={pattern=north west lines,pattern color = brown}}
			\addlegendentry{Tit-for-tat};
			\addlegendimage{color=teal,fill=teal!30,postaction={pattern=north east lines,pattern color = teal}}
			\addlegendentry{Nice Tit-for-tat};
			\end{axis}
			\end{tikzpicture}
		}
		\hfill
		\caption{Average utility of negotiating agents}
		\label{results}
	\end{figure*}
	The time taken by our agent for each round has been set empirically. When our agent makes a choice among 200 bids, there is generally one that results into a high utility both for it and its opponent. In order to explore the tree in depth, we let it enough time to generate about 50'000 simulations. Then, to make a proposal, our agent takes about 3 minutes. This duration meets the expectation of the real-world applications of our industrial partners. From these values, we get $\alpha=0.489$.
	
	\subsection{Opponent's Description}
	In this section, we provide a description of the confronted agents. It is impossible to compare our agents with state-of-the-art ones, as they all rely on the supposition that there is a publicly known deadline. The only three agents able to negotiate without a given deadline are the RandomWalker, the Tit-for-Tat agent and its more evolved variant, the Nice Tit-for-Tat.
	\subsubsection{RandomWalker}
	The RandomWalker is described by Baarslag in \cite{Baarslag2013} and makes random proposals.
	
	\subsubsection{Tit-for-Tat}
	Tit-for-tat agent was first described by Faratin \textsl{et al.} in \cite{Faratin1998Negotiation}. This agent makes a concession whenever its opponent makes concessions itself. Several possible implementations are given. Here, when it has received fewer than 2 proposals the agent makes the most interesting proposal from its own perspective (it generates 10'000 random proposals and chooses the best one for its utility function). For the other proposals, the agent looks at the last two proposals of its opponent and computes the made concession. This concession may be positive or negative. It adds this concession to the utility of its own last proposal. It then searches for a proposal with the closest utility to this target value. In our implementation, 10'000 proposals are generated to this end.

	\subsubsection{Nice Tit-for-Tat}
	Nice Tit-for-Tat is somehow, a more evolved version of the Tit-for-Tat. It has been described by Baarslag \textsl{et al.} in \cite{Baarslag2013tit}. The goal of this work is to comply with domains where mutual agreement is possible. The agent then uses the same opponent utility modeling as our agent and uses it to estimate the Nash point of the setting, \textsl{i.e.} the agreement maximizing the product of the utilities of the agents. The concession rate is computed between the first and the last bids of the opponent as the percentile of the distance between its first bid and the Nash point. The corresponding concession is made from the agent point of view. Among the equivalent offers, the utility model is also used in order to choose the best bid for the opponent among equivalent ones for the agent. The only difference with the version used in our experiments and the agent proposed in \cite{Baarslag2013tit} is the acceptance strategy. Indeed, the strategy proposed in the original version which is presented in \cite{Baarslag2013} depends on the deadline of the negotiation in order to take the time pressure into account. Here, we provide the Nice Tit-for-Tat with a simplified version of its acceptance strategy, which corresponds to the same acceptance strategy as our agent.
	
	\subsection{Experimental Protocol}
	\textsc{Genius} makes it possible to negotiate on numerical or categorical issues, but not yet on continuous ones. In order to evaluate our agent, we want to target a negotiation domain that is at the same time neutral enough to show the generic aspects of our work, but complex enough to motivate its use. ANAC is an international competition used to determine the effective negotiation strategies. The negotiation domain used in ANAC~2014 \cite{Fukuta2016} fits well to this objective. Subsequent competitions focused on multilateral negotiation and specific application domains. In ANAC~2014 domain, issues are numerical, varying from 1 to 10. Several domains have been proposed, varying from 10 to 50 issues. In order to reduce computational complexity which is not the concern of this work, we use the 10-issue version. The utility functions are non-linear, and the reserve utility is set to 0, which is the minimal outcome value for the agents. As the time pressure is supposed to vary over time, we do not use a discount rate. While it does not exactly correspond to the targeted application, its complexity ($10^10$ possible proposals) makes a suitable test bench for our negotiation strategy. In order to simulate the fact that the time pressure is unknown to the agent, we put a very large deadline, so that it is never reached.
	
	\subsection{Results}
	\Cref{results} displays the utility of the agents when negotiating with each other using a histogram. The results are averaged over 20 negotiation sessions with each profile, with error bars representing the standard deviation from the average.
	
	Note that the two preference profiles are very different from each other, and not symmetrical at all. This specificity explains the fact that for all the agents in all configurations, utility is always higher with Profile~2 than it is with Profile~1.
	
	As shown on \Cref{res-rw}, our agent is able to beat the Random Walker in every situation, even when its preference profile is Profile~1 and Random Walker's is Profile~2. It is interesting to note that the negotiations with Random Walker are very short with only 3.1~proposals in average: 2.5 when it gets Profile~1 and 3.7 proposals when our agent gets Profile~2. This difference can be explained by the fact that it is easy to find agreements with very high results for Profile~1 (0.9 or more) and high results for Profile~2 (0.6). In most of the negotiations, the first proposal of our agent is of this kind. In that case, Random Walker is more likely to generate a proposal with utility lower than the one proposed by our agent and accepts it, generating a utility of 0.6 for Random Walker and a negotiation session consisting of a single proposal.
	
	Negotiating with Tit-for-Tat is harder, as we can see on \Cref{res-tft}. Our agent gets a lower utility than the Random Walker but is able to beat Tit-for-Tat. The expectation level of Tit-for-Tat also generates much longer negotiations: 34.2 proposals on average with 31.55 proposals when it gets Profile~1 and 36.85 proposals when our agent gets Profile~2. This result can be explained the same way as the results of the negotiation with Random Walker.
	
	The negotiations with Nice Tit-for-Tat never ends: the agents keep negotiating forever. Our MCTS-based method refuses to make a concession significant enough to have a chance to be accepted by Nice Tit-for-Tat, considering the high expectation it has by using  the Nash point. Reciprocally, Nice Tit-for-Tat, without time pressure, and dynamic adaptation of its acceptance strategy, does not accept the proposals of our agent. By looking at the internal state of the Nice Tit-for-Tat, we also see that its estimation of the Nash point is incorrect: it expects a utility above the real one.
	
	We propose instead an indirect evaluation by confronting Nice Tit-for-Tat with Random Walker, in the same setting. The results are represented on \Cref{res-ntft}. The performances of both agents are comparable, considering the standard deviation of the series.
	
	\section{Conclusion}
	\label{conclusion}
In this paper, we presented an automated negotiation agent able to negotiate in a context where agents do not have predetermined deadline, neither in time nor in rounds, and where the negotiation domain can be composed of numerical, continuous and categorical issues. We described this setting as an extensive game and described a negotiation strategy based on a specific implementation of MCTS relying on two opponent models and a pruning strategy. One of them is the Gaussian process regression, which relies on a covariance function. We tested several covariance functions and chose the one that provides better results in context similar to ours. 

Experiments were run in the context of a large negotiation domain, with nonlinear utility functions using different preference profiles.  The experimental results are promising: against  all the agents that can negotiate in its negotiation  domain, our agent outperformed Random Walker and Tit-for-Tat and draws with Nice Tit-for-Tat. This work therefore indicates that techniques from games such as MCTS can be used with success in automated negotiation. However, the modularity of the architecture and the variety of strategies proposed on General Game Playing  and Machine Learning areas are a huge advantage when it comes to optimizing the agent for a specific application domain.

Among the perspectives of this work, we would like to create a customized version of our agent and adapt it to the context where the deadline is known, in order to make it available for these applications. Our agent can already be used in this context, but the fact that it does not exploit this information may make it less efficient than its opponents. Another possible direction would be to adapt it to the multilateral context. In fact, there would be little modification to make it available for a context of stacked alternate protocol or a many-to-many bargaining scenario, since MCTS has already been used in n-player games. The use of our agent in its industrial context, in particular with corresponding negotiation domains would yield very interesting results. Last, we would like to improve our agent by using MCTS variations. It would be interesting for instance to test other kinds of pruning. The use of traditional MCTS techniques such as Rapid Action Value Estimation or All Moves As First to reduce the number of simulations while keeping their intrinsic qualities would be interesting.
 \bibliographystyle{splncs04}
	\bibliography{biblio.bib}
	\end{document}